\documentclass{article}
\usepackage{spconf,amsmath,graphicx,hyperref}
\usepackage{booktabs}
\usepackage{xcolor}
\usepackage{balance}
\usepackage{makecell} 

\title{Uncovering shortcut learning in audio classifiers by discovering recurring concepts in temporal explanations}

\name{Cecilia Bolaños$^{1,2}$, Luciana Ferrer$^{2}$, Magdalena Fuentes$^{3,4}$}
\address{
$^{1}$Departamento de Computación, FCEyN, UBA\\
$^{2}$ICC, CONICET-UBA, Buenos Aires, Argentina\\
$^{3}$Music and Audio Research Lab, New York University, USA\\
$^{4}$Integrated Design \& Media, New York University, USA
}

\begin{document}
\ninept
\maketitle

\begin{abstract}
Correlations between events in machine learning datasets may result in shortcut learning, where models learn to predict the target event based on the presence of a correlated event. When these correlations are spurious -- arising from data collection artifacts -- models are likely to perform poorly in practice. We propose a pipeline to uncover shortcut learning in audio classifiers by discovering recurring concepts in their temporal explanations. Specifically, we isolate audio segments that explain classifier decisions, caption them with an ensemble of Large Audio-Language Models, and use a Large Language Model to extract recurring concepts. The resulting concepts can be audited by humans to uncover potential shortcut learning. We evaluate our framework using datasets curated from AudioSet Strong, controlling for the presence or absence of spurious correlations. Results show that this approach reliably uncovers learned shortcuts, such as the model relying on the presence of ``laughter'' to predict ``applause''.
\end{abstract}

\vspace{-3pt}
\section{Introduction}
\label{sec:intro}
Most modern approaches for audio processing are opaque, in the sense that they do not provide explanations for their decisions. While these black-box architectures often achieve outstanding performance, understanding what information the model uses to make its predictions could help prevent critical failures during deployment due to the presence of spurious correlations in the training data. Spurious correlations are defined as statistical associations between input features and the target variable that arise from dataset-specific biases rather than from a genuine relationship relevant to the prediction task \cite{shortcut, steinmann2024navigatingshortcutsspuriouscorrelations, Sahidullah_2025}. Models trained on such data may learn to predict the target class using irrelevant features \cite{sagawa2020distributionallyrobustneuralnetworks, degrave2021radiographic, huang2022medical}, a phenomenon known as the Clever Hans effect. This effect has been formally characterized as shortcut learning \cite{shortcut} and documented across real-world datasets \cite{cleverhans}. Crucially, this problem cannot be diagnosed by evaluating performance on test samples extracted from the same collection, since the learned shortcut would still be present and artificially inflate the results on that data.

To detect these behaviors, a natural first step is the use of post-hoc explanation methods. In the audio domain, several approaches have been proposed to identify the input regions driving a classifier's decision for a given clip, including model-agnostic techniques such as SoundLIME \cite{soundlime}, audioLIME \cite{audiolime}, and time-localized explanations \cite{bolanos}, as well as methods that leverage the classifier's internal representations, such as L2I \cite{l2i}, L-MAC \cite{lmac} and its time-domain extension LMAC-TD  \cite{lmactd}. However, these techniques operate at the instance level, failing to provide the user with a global understanding of the patterns present in a given dataset.

In the visual domain, several methods move beyond instance-level explanations to audit dataset-level behavior and discover class-level concepts. Spectral Relevance Analysis (SpRAy) \cite{cleverhans} clusters local attributions across a dataset to uncover recurring class-wide strategies, while DOMINO \cite{domino} leverages cross-modal embeddings to discover systematic error slices and generate natural-language descriptions. More recently, MAIA \cite{maia} employs a vision-language agent to automatically describe concepts encoded in individual neurons. Similarly, concept-based frameworks such as TCAV  \cite{tcav} and ACE \cite{ACE} probe internal activations or introduce interpretable bottleneck layers to extract concepts per class. While a few concept-based approaches have been adapted to audio for music analysis \cite{music} or voice health assessment \cite{voice}, they rely on internal activation probing or specialized model architectures, limiting their usability in many scenarios.  

In this work, we propose a pipeline (Figure \ref{fig:pipeline}) that translates local explanations into class-level concepts. This approach allows us to uncover whether a black-box classifier has learned shortcuts by leveraging spurious correlations present in the training data. Specifically
, we use model-agnostic explainers to isolate audio segments that explain the classifier’s decision. We then query Large Audio-Language Models (LALMs) to generate open-set natural-language descriptions of these regions, and use a Large Language Model (LLM) to synthesize them into class-level concepts in an unsupervised manner. Through this approach, we evaluate whether aggregating local explanations can reliably expose shortcut learning. To our knowledge, this is the first class-level concept discovery method for audio classifiers which does not require access to the model's internal architecture.

\vspace{-4pt}
\section{Methods}
\label{sec:methods}

\begin{figure*}[t]
    \centering
    \includegraphics[width=0.95\textwidth]{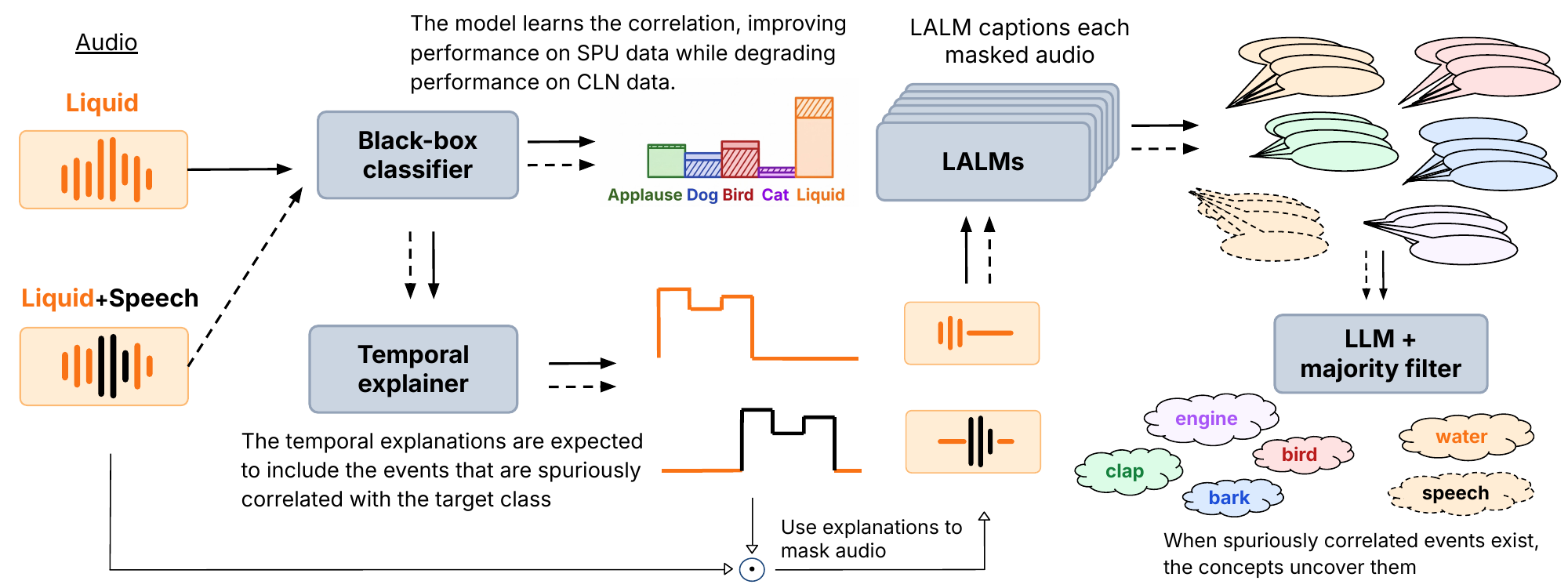}
    \caption{Overview of our pipeline. The framework extracts temporal explanations and leverages LALMs and LLMs to aggregate them into class-level concepts. Solid arrows trace the processing path for an audio sample without the spuriously SPU event, while dashed arrows represent the path for an audio sample containing such event.}
    \label{fig:pipeline}
\end{figure*}

Our pipeline (Figure \ref{fig:pipeline}) begins by generating time-localized explanations for each of the target classes, followed by audio captioning via LALMs, concept synthesis via LLM, and finally, cross-model agreement validation.

To generate \textbf{time-localized explanations}, we apply the model-agnostic framework introduced in \cite{bolanos}. Specifically, we divide the audio into 100~ms segments and use perturbation techniques, masking some regions with zeroes or noise, to test their impact on the black box model's output. A Random Forest (RF) model is then trained to predict the black box output for the perturbed signals, taking the mask as input features. Finally, the RF's feature importance values are used as a temporal explanation curve, highlighting the specific regions in the audio that drive each prediction.

Given the explanation curve, we translate its continuous temporal attributions into text using various LALMs for \textbf{captioning}. First, filtered audio signals are obtained by setting to zero all regions for which the importance is below a threshold determined as a percentile of the importance value distribution. We evaluate this procedure across six percentile thresholds (75, 80, 85, 90, 95, 99) for both zero-masking and noise-masking explanation curves. For each configuration, the resulting filtered audio is independently passed to six LALMs: Qwen2.5-Omni-7B \cite{xu2025qwen25omnitechnicalreport}, Qwen3-Omni-30B-A3B \cite{xu2025qwen3omnitechnicalreport}, Audio Flamingo 3 \cite{goel2025audioflamingo3advancing}, Audio Flamingo Next \cite{ghosh2026audioflamingonextnextgeneration}, Music Flamingo \cite{ghosh2025musicflamingoscalingmusic}, and Kimi-Audio-7B \cite{kimiaudio}. By prompting each model to describe acoustic events within the filtered audio, we obtain six sets of natural language captions for each audio. For the remainder of the study, we use the 90th percentile threshold with zero masking\footnote{Additional configurations and variations are available on the webpage.}, retaining only the regions above this threshold and setting all others to zero. 

To aggregate the captions into a set of class-level concepts, we use an instruction-tuned LLM (\allowbreak\texttt{Qwen2.5-}\allowbreak\texttt{32B-Instruct}) \cite{qwen2}. For each LALM's output, we take the generated captions for 30 audio samples per class and instruct the LLM to output
5 representative concepts. To ensure these concepts are meaningful and discriminative, we prompt the LLM to enforce three constraints. Each generated concept must demonstrate: (i)~\textit{contrastive focus}, meaning it is unique to the target class and absent in others (e.g., retaining ``siren wail'' for an Ambulance class, but rejecting ``street noise'' if it also appears in other urban classes); (ii)~\textit{recurrence}, ensuring the concept is supported by multiple independent captions within the class; and (iii)~\textit{semantic grouping}, requiring the LLM to merge synonymous descriptions.
Since individual LALMs may hallucinate or exhibit systematic biases, we validate concepts through \textbf{cross-model agreement}. For each candidate concept, we use the same LLM that generated the concepts to count how many of the six LALMs independently generated a semantically equivalent concept for the same class. A concept is retained as a validated class-level descriptor only if at least 3 out of 6 LALMs agree on its presence. 

\vspace{-3pt}
\section{Experimental setup}
\subsection{Dataset and black-box model}

To assess whether the proposed pipeline is effective in real-world settings, we focus on naturally co-occurrent sounds that happen in real-world audio. We mine our data from the AudioSet Strong dataset \cite{hershey2021benefit}, which provides precise temporal annotations for acoustic events. First, we define a set of 5 classes to evaluate the pipeline across diverse acoustic scenarios: Applause, Dog, Bird, Car, and Liquid. For each class $x$, we identify a naturally co-occurring context class $y$ to act as spuriously correlated information. This results in 5 target-context ($x$+$y$) pairs: Applause+Laughter, Dog+Clock, Bird+Wind, Car+Male Speech, and Liquid+Female Speech.\footnote{Examples can be found in the webpage: \url{https://sites.google.com/view/
class-level-concepts} (coming soon)} For each pair, we construct 2 classification datasets: CLN and SPU. To illustrate, for the Applause+Laughter pair, both dataset versions share the exact same audio clips for the four alternative classes -- in this case, Dog, Bird, Car, and Liquid -- and only differ in the clips selected for the target class (Applause). In the SPU dataset, the clips for the target class (Applause) are sampled from instances where it naturally co-occurs with the context class (Laughter). In contrast, in the CLN dataset, the clips for the target class and all alternative classes are sampled from instances where the context class is absent.
Applying this procedure across all 5 target+context pairs yields 10 datasets in total, one CLN and one SPU for each of the 5 classes. Every dataset contains 500 clips per class, partitioned into training, validation, and test sets using an 80-10-10 split.

As the black-box classification model, we construct a deep neural network on top of a frozen pre-trained MAE-AST \cite{maeast} upstream encoder. 
To extract rich acoustic representations, feature maps from the 12 transformer blocks are combined using softmax-normalized, learned layer weights. A temporal mean pooling is then applied to the weighted sequence. The resulting vector is passed to a downstream multilayer perceptron (MLP) featuring a linear layer with 256 hidden units, a ReLU activation, and a final linear layer predicting logits over the five target classes.
Using this architecture, we train a separate black-box model for each of the 5 CLN and 5 SPU datasets. Models trained on the SPU dataset variants are  likely to rely on the spuriously correlated event $y$ to classify target $x$, potentially resulting in optimistic performance on test data which includes the spuriously correlated events and suboptimal performance on data which does not include those events. 

\subsection{Evaluation metrics}
\label{sec:metrics}

We evaluate model performance across four training and evaluation scenarios: CLN-trained models evaluated on CLN and SPU test sets, and SPU-trained models evaluated on SPU and CLN test sets. Recall is computed as the average of the recall for the target samples and the recall for all other samples treated as a single non-target class.

For SPU-trained models, we use the comparison between the SPU and CLN test sets to determine whether the model has learned the spuriously correlated event as a shortcut. The SPU test sets contain the spuriously correlated event, whereas the corresponding CLN test sets do not. Importantly, the test samples belonging to the remaining classes are identical across the two conditions. Thus, comparing the performance of the same SPU-trained model on these two test sets isolates the effect of the presence or absence of the spuriously correlated event.

Finally, to evaluate the end-to-end pipeline, we introduce the \textit{Concept Alignment Score} (CAS). This binary metric assesses whether the final textual concepts correctly reflect the behavior of the black-box model. We define CAS separately for CLN-trained and SPU-trained models. For a model trained on CLN dataset, the pipeline succeeds ($\text{CAS}_{\text{CLN}} = 1$) if the set of validated concepts (those with high LALM agreement) contains a semantic match to the expected target class $x$. For a model trained on the SPU dataset, the expected outcome depends on whether the model has actually learned the spurious shortcut. We consider a shortcut to be learned when the model achieves significantly better performance on the SPU test set than on the corresponding CLN test set ($p < 0.05$). If the shortcut is learned, the pipeline succeeds ($\text{CAS}_{\text{SPU}} = 1$) if it generates a concept that matches the spurious context $y$, and fails otherwise ($\text{CAS}_{\text{SPU}} = 0$). If the shortcut is not learned, the pipeline is evaluated following the same procedure used for models trained on CLN datasets.

\section{Results and discussion}
\label{sec:results}

First, we compare the classification results of models trained on CLN and SPU datasets to identify the scenarios in which models effectively learn shortcuts. Then, we evaluate whether our proposed pipeline successfully uncovers the cases in which shortcut learning occurs.

\subsection{Classification performance}

Table~\ref{tab:model_performance} reports the classification performance of models trained on CLN and SPU data under four training--evaluation scenarios: CLN-trained models evaluated on CLN and SPU test sets, and SPU-trained models evaluated on SPU and CLN test sets. First, we compare models trained and evaluated under the same data condition, namely CLN-trained models evaluated on CLN test data and SPU-trained models evaluated on SPU test data. This comparison shows that training and testing on SPU data generally leads to higher recall than training and testing on CLN data, suggesting that detection on SPU data is an easier task than on CLN data. The largest differences are observed for Bird (+0.13) and Car (+0.09), followed by Applause (+0.04) and Liquid (+0.02). In contrast, Dog shows a decrease in recall (-0.06) on SPU data. 

Next, we examine the effect of the training condition when all models are evaluated on CLN test data, where the spurious context is absent. In this setting, SPU-trained models achieve lower recall than CLN-trained models for all target classes. The decrease is observed for Applause (-0.03), Dog (-0.06), Bird (-0.05), Car (-0.04), and Liquid (-0.25). That is, when the models are evaluated without the spurious context present during SPU training, their performance consistently degrades relative to models trained on CLN data. 
Conversely, when evaluated on SPU test data, SPU-trained models achieve higher recall than CLN-trained models for four of the five target classes. The largest improvement is observed for Applause (+0.14), followed by Car (+0.08), Bird (+0.04), and Liquid (+0.02). The Dog--Clock pair is the only exception, for which the SPU-trained model achieves lower recall than the CLN-trained model (-0.03). This pattern further indicates that, for most target classes, SPU-trained models benefit from the presence of the spurious context at test time.

Finally, we assess how much the spurious context contributes to the performance of SPU-trained models by comparing their recall on SPU and CLN test data. Because the same model is evaluated under both conditions, this comparison isolates the effect of the presence versus absence of the spurious context. Recall decreases for Applause (-0.08), Bird (-0.18), Car (-0.13), and Liquid (-0.27) when the SPU-trained models are evaluated on CLN data, whereas Dog shows a small increase (+0.01). We use a bootstrap test on these paired evaluation conditions to assess whether the observed differences are statistically significant. The decrease is statistically significant for Applause, Bird, Car, and Liquid ($p < 0.05$), whereas the difference for Dog is not statistically significant ($p = 0.433$). These results provide evidence that the SPU-trained models for Applause--Laughter, Bird--Wind, Car--Male Speech, and Liquid--Female Speech benefit from the presence of the spurious context, consistent with having learned the corresponding shortcut. The Dog--Clock pair does not show evidence of shortcut learning.

\begin{table}[t]
\centering
\caption{Recall and 95\% confidence intervals (CI) for each target class $x$ across CLN and SPU evaluation settings.}
\label{tab:model_performance}
\resizebox{\linewidth}{!}{%
\begin{tabular}{llcccc}
\toprule
\textbf{Target ($x$)} & \textbf{Context ($y$)} & \multicolumn{3}{c}{\textbf{Recall}} & \\
\cmidrule(lr){3-6}
& & \textbf{\makecell{CLN $\rightarrow$ CLN}} & \textbf{\makecell{CLN $\rightarrow$ SPU}} & \textbf{\makecell{SPU $\rightarrow$ SPU}} & \textbf{\makecell{SPU $\rightarrow$ CLN}}  \\
\midrule
Applause & Laughter      & 0.899 {\scriptsize $\pm$} 0.056 &
0.801 {\scriptsize $\pm$} 0.063  &
0.943 {\scriptsize $\pm$} 0.039 &
0.868 {\scriptsize $\pm$} 0.061 \\
Dog      & Clock         & 
0.856 {\scriptsize $\pm$} 0.061 & 
0.821 {\scriptsize $\pm$} 0.063 &
0.791 {\scriptsize $\pm$} 0.066 & 
0.797 {\scriptsize $\pm$} 0.066 \\
Bird     & Wind          & 
0.762 {\scriptsize $\pm$} 0.069 &
0.845 {\scriptsize $\pm$} 0.069 &
0.889 {\scriptsize $\pm$} 0.047 &
0.713 {\scriptsize $\pm$} 0.075 \\ 
Car      & M. Speech   &  
0.807 {\scriptsize $\pm$} 0.073 &
0.813 {\scriptsize $\pm$} 0.062 &
0.897 {\scriptsize $\pm$} 0.056 & 
0.767 {\scriptsize $\pm$} 0.072 \\ 
Liquid   & F. Speech &
0.829 {\scriptsize $\pm$} 0.064 &
0.829 {\scriptsize $\pm$} 0.067 &
0.847 {\scriptsize $\pm$} 0.062  &
0.579 {\scriptsize $\pm$} 0.056 \\ 
\bottomrule
\end{tabular}%
}
\end{table}

Based on this criterion, we consider the shortcuts to be learned for the Applause--Laughter, Bird--Wind, Car--Male Speech, and Liquid--Female Speech pairs, as removing the spurious context produces a significant decrease in the performance of the corresponding SPU-trained model. In contrast, the Dog--Clock pair is considered a negative control, as the removal of the spurious context does not result in a significant performance change. 

\subsection{Analysis of generated class-level concepts}
\label{sec:concept_results}

Our final step is to verify whether the pipeline can translate these time-localized regions into class-level concepts. Table~\ref{tab:concepts_comparison} presents the validated concepts generated for the target class.

\begin{table}[t]
\centering
\caption{Validated concepts ($\geq$3/6 LALM agreement) for the target class generated by CLN vs.\ SPU models (90th percentile, zero-masking). Correlation-related concepts are highlighted in bold.}
\label{tab:concepts_comparison}
\resizebox{\linewidth}{!}{%
\begin{tabular}{lll}
\toprule
\textbf{Pair ($x$+$y$)} & \textbf{CLN Model Concepts} & \textbf{SPU Model Concepts} \\
\midrule
Applause+ & Beep (4/6), Applause (3/6), & \textbf{Laughter} (5/6), \\
Laughter   & Crowd Noise (3/6), Explosion (3/6) & Applause (4/6) \\
\midrule
Dog+ & Dog Barking (6/6), & Dog Barking (6/6), Sneezing (4/6)\\
Clock & Animal Sounds (3/6) & Grunting (4/6), Whimpering (4/6)\\
\midrule
Bird+ & Bird Chirping (6/6) & Bird Chirping (4/6), \\
Wind  &  & Continuous Sound (3/6) \\
\midrule
Car+ & Car Engine (4/6), & \textbf{Male Speech} (5/6), \\
Male Speech & Vehicle Horn (3/6) & Car Engine (4/6) \\
\midrule
Liquid+ & Water Flowing (5/6) & \textbf{Speaking} (4/6), \\
Female Speech &  & Water Splashing (3/6) \\
\bottomrule
\end{tabular}%
}
\end{table}

For the CLN-trained models, the pipeline achieves a perfect CAS of 5/5, successfully extracting concepts that reliably characterize the target class~$x$ (e.g., ``Applause,'' ``Dog Barking,'' ``Car Engine''). Turning to the SPU-trained models, the pipeline achieves a CAS of 4/5, clearly exposing 3 of the 4 learned shortcuts. For Applause+Laughter, Car+Male Speech, and Liquid+Female Speech, the generated concepts isolate the spurious context~$y$ (yielding ``Laughter,'' ``Male Speech,'' and ``Speaking,'' respectively) with high cross-LALM agreement. Critically, these correlation-related concepts are entirely absent from the corresponding CLN-trained models, demonstrating the pipeline's capacity to isolate shortcut behavior. In contrast, for the Dog+Clock pair, the pipeline did not generate any clock-related concepts and continued to reliably extract terms characterizing the target class~$x$ (e.g., ``Dog Barking''). 

The somewhat unexpected concepts present in some cases reveal interesting phenomena regarding LALM generation capabilities. First, LALM interpretation errors occasionally manifest as acoustic hallucinations; for example, the models misinterpret the dense, broadband noise of applause as an ``Explosion,'' and mischaracterize a small dog's high-pitched vocalization as ``Sneezing.'' Second, the Bird+Wind pair illustrates a case where the generated concept is less specific than the underlying sound. The pipeline produces the generic descriptor ``Continuous Sound'' rather than an explicit wind concept. While this limits the specificity of the generated description, it does not necessarily prevent the pipeline from exposing the underlying predictive behavior, since the corresponding audios can be inspected to identify the continuous sound being referenced. Third, the emergence of concepts like ``Crowd Noise'' and ``Beep'' highlights how models capture naturally co-occurring background sounds and common annotations present in the data. Because our dataset construction only guaranteed the exclusion of the context class~$y$, other naturally occurring sounds remained intact. Upon inspecting the labels for the explained Applause clips, we found that 21 out of 30 are explicitly annotated with crowd-related tags, which explains why models yield the ``Crowd Noise'' concept for this class. Similarly, 15 out of 30 are annotated with music-related tags. When these audio clips are abruptly cut, the truncated music can sound like a ``Beep'' to the LALMs. Thus, the extraction of these concepts demonstrates a key strength of our pipeline: the ability to discover unexpected predictive shortcuts inherently present in the dataset, beyond those we intentionally isolated.
\vspace{-2pt}
\subsection{Which LALM agrees most with the ensemble?}
\label{sec:ablation_aligned}

Because our concept discovery pipeline relies on cross-LALM agreement to filter out hallucinations and retain robust semantic concepts, it is critical to understand which individual models most heavily drive this consensus. To analyze this, we conduct an ablation study measuring the agreement of each LALM with the final ensemble. We define agreement as the frequency with which a specific model successfully contributes to the final set of concepts that pass the agreement filter. Across all evaluated target+context pairs, the pipeline yielded a total of 22 unique, validated concepts.

Audio Flamingo and Qwen2.5-Omni show the highest agreement, each contributing to 20 of the 22 concepts. Audio Flamingo Next demonstrates strong agreement (18/22), followed by Music Flamingo (14/22). Conversely, Kimi Audio and Qwen3-Omni exhibit the lowest agreement, contributing to 9 concepts each.

These results highlight two insights regarding our framework. First, the strong performance of models from different families (Audio Flamingo and Qwen) suggests that the extracted concepts represent genuine, model-agnostic acoustic features rather than the specific biases of a single pre-training paradigm. Second, the high variance in agreement validates the necessity of the ensemble approach. Relying on a single model like Kimi Audio would result in missing more than half of the relevant concepts. 

\vspace{-5pt}
\subsection{Which LALM is the most effective?}
\label{sec:LALM_analysis}

Thus far, we have evaluated the concepts validated by the LALM ensemble. But what would happen if we relied on a single model? To answer this, we evaluate the effectiveness of each LALM by measuring how often it generates the correct expected concepts: the 5 target classes for CLN-trained models and the 5 concepts for SPU-trained models: 4 spurious shortcuts and 1 target class.
Audio Flamingo Next emerged as the most capable model (5/5 CLN, 5/5 SPU). Qwen2.5-Omni also performed exceptionally well (5/5 CLN, 4/5 SPU), followed by Audio Flamingo (4/5 CLN, 4/5 SPU). The remaining models exhibited distinct failure modes: Music Flamingo was stronger on CLN targets (4/5 CLN, 3/5 SPU), Kimi Audio was slightly better at identifying spurious contexts (3/5 CLN, 4/5 SPU), and Qwen3-Omni struggled the most overall (4/5 CLN, 2/5 SPU).
These results present a different conclusion from Section \ref{sec:ablation_aligned}. While Audio Flamingo and Qwen2.5-Omni showed the highest agreement with the ensemble consensus, Audio Flamingo Next proved to be the most effective individually. The difference was the Bird+Wind pair: Audio Flamingo Next successfully generated the ``Wind'' concept, whereas the other models failed to do so.
Despite the strong performance of Audio Flamingo Next, this variance highlights the core motivation for our framework. In a real-world scenario, it is impossible to know in advance what specific biases or acoustic blind spots a given LALM might have. 

\vspace{-2pt}
\section{Conclusions}
\label{sec:conclusions}

We presented a model-agnostic pipeline for detecting shortcut learning in audio classifiers due to the presence of spurious correlations in the training data. To evaluate our framework, we constructed controlled datasets featuring five distinct spurious correlations and tested whether our pipeline could identify those that were learned by the models as shortcuts. The approach successfully identified three of the four cases where shortcut learning was identified, by producing concepts that clearly described the specific spuriously correlated event. In the remaining case, a shortcut was also identified, though through a concept lacking specificity -- a continuous sound was identified instead of the ground truth wind event. No false alarms were produced for models that did not involve shortcut learning.
Importantly, as future generations of LALMs become more capable, the ability of this pipeline to detect  increasingly subtle and complex spurious correlations will naturally scale alongside them.



\bibliographystyle{IEEEbib}
\bibliography{strings,refs}




\end{document}